\def\be{\begin{equation}}
\def\ee{\end{equation}}
\def\bea{\begin{eqnarray}}
\def\eea{\end{eqnarray}}
\def\ba{\begin{align*}}
\def\ea{\end{align*}}

\documentclass[aps,showpacs,showkeys,twocolumn,amsmath,10pt,superscriptaddress,floatfix,nofootinbib]{revtex4-1}
\usepackage{amsfonts,amssymb,latexsym,amsmath,epsfig,bm,times}
\usepackage{color}
\usepackage{subfigure}
\usepackage{bm}
\usepackage{graphicx}
\begin{document}
\title{Controlled Spin Transport in Planar Systems Through Topological Exciton}
\author{Kumar Abhinav}
\email{kumarabhinav@iiserkol.ac.in}
\affiliation {Indian Institute of Science Education and Research Kolkata, Mohanpur-741246, West Bengal, India}
\author{Prasanta K. Panigrahi}
\email{pprasanta@iiserkol.ac.in}
\affiliation {Indian Institute of Science Education and Research Kolkata, Mohanpur-741246, West Bengal, India}
\date{\today}
\begin{abstract}
It is shown that a charge-neutral spin-1 exciton, realizable only in planar systems like graphene,
can effectively be used for controlled spin transport in such media. The excitonic 
bound state is destabilized by quantum fluctuations, characterized by a threshold for excitation and melts
in a smooth manner under thermal fluctuations. This planar exciton
differs from the conventional ones, as it owes its existence to the topological Chern-Simons (CS) term. The parity
and time-reversal violating CS term can arise from quantum effects in systems with parity-breaking mass-gap. The 
spinning exciton naturally couples to magnetic field, leading to the possibility of controlled spin transport.
Being neutral, it is immune to adverse effects, afflicting spin transport by charged fermions.
\end{abstract}
\maketitle

Lower dimensional systems exhibit a host of unique features \cite{N0}, absent in their higher-dimensional counterparts. 
Realization of Majorana fermion \cite{N1}, thin film topological insulators (TIs) \cite{N2}, charge density
wave \cite{N3}, helical transport \cite{N4} and charge-spin separation \cite{N5} characterize one-dimensional
systems. Planar (2+1 dimensional) materials have led to physical realization of large spin-orbit coupling \cite{N6}, 
spin-momentum locking \cite{N7} and low-energy Dirac fermions \cite{Gra3} etc. 2+1 dimensional
systems are endowed with non-trivial inherent topology, which manifests in quantum Hall
effect \cite{Thouless,QH2,Wen1,Stone1} and boundary conductivity in TIs \cite{TI,TI0,TI2}, having symmetry
protected topological order \cite{N9}. 
\paragraph*{}The characteristic non-trivial feature of the planar world, absent in their even-dimensional
counterparts, is the manifestation of the topological Chern-Simons (CS) term,

\be
\mathcal{L}_{CS}=\frac{\bar{\mu}}{2}\epsilon^{\mu\nu\rho}a_{\mu}\partial_{\nu}a_{\rho}.\label{E1}
\ee
in the gauge sector, due to which, propagating photons can have gauge-invariant mass \cite{TopM1,TopM2,FA2}. Interacting
particles can acquire additional spin \cite{FA0,FA2,H,PKP01,TopM2}, leading to a change in their statistics \cite{Myrheim,FA0,H2,FA1,KoSe,DasPanda}. The CS term
arises due to quantum effects, in theories with parity-breaking massive fermions \cite{TopM1,TopM2,Red,Boy,DM}
and bosons \cite{PKP1}. The advent of planar materials like graphene \cite{Gra1,Gra2,Gra3} and TIs \cite{TI0,TI} have led to
physical realization of low-energy `relativistic' planar systems. Massive fermions have been experimentally realized by
inducing sub-lattice density asymmetry in bi-layer graphene under transverse electric field \cite{MM1,GraM02} and in
graphene mono-layer, misaligned with the substrate \cite{GraM2}. The hexagonal structure of TIs leads to a single Dirac point, with
locally inducible gap through magnetic induction \cite{TEM,TI,TEM1,TIM1}. The importance of mass term lies in the fact that 
in 2+1 dimensions, mass-less particles are spin-less. Furthermore, parity-breaking mass term is essential for generating
the CS term, not present in natural electromagnetism.
\paragraph*{}In this letter, we demonstrate controlled spin transport in graphene-like materials, through charge-less
spin-1 excitons, owing their existence to this topological CS term. The fact that, the in-plane optical phonon and
lattice curvature couple as $U(1)$ gauge fields to the emergent Dirac fermions of graphene, leads to possible realization of the topological CS term, along with a dynamic gauge part, through quantum corrections. The quantum fluctuations destabilize the
bound state, at zero temperature. This leads to a parametric threshold for excitation of this topological
quasi-particle. The bound-state smoothly melts at high temperature.
\paragraph*{}Quite some time back, Hagen demonstrated the existence of this spin-1 {\it exciton}, a weakly
bound fermion-antifermion pair, arising due to the CS term in 2+1 dimensional gauge theory \cite{H}. This is unlike the
non-topological excitons, recently seen in graphene nano-ribbons \cite{Ge5} and in TIs \cite{TE2}. Being a
spin-1 object, this exciton naturally couples to magnetic field \cite{Kogan1}, leading to possible controlled {\it spin-transport}
in planar systems like graphene, which is fundamentally different from the spin-transport of spin-$\frac{1}{2}$
charged excitations, already realized in graphene \cite{ST01} and in TIs \cite{ST02}. Similar fermionic bound
states, possessing different spins, naturally manifest in various branches of physics. The Nambu--Jona-Lasinio
model \cite{NJL} realizes pseudoscalar pions as mass-less bound state of protons and neutrons. The space-time
parity-breaking low-energy topological Wess-Zumino-Novikov-Witten term \cite{WZNW} correctly describes the
interactions of these particles in 3+1 dimensions. 
\paragraph*{}As mentioned earlier, in graphene, in-plane optical phonons \cite{Neto,Ando1} and defect-induced
lattice curvature \cite{Neto,GFg} couple to the low-energy relativistic fermions as $U(1)$ gauge fields, the
former having {\it spatial} components $\left(a_x,a_y\right)\propto\left(u_y,-u_x\right)$ \cite{Ando1}, with
$u_{x,y}$ depicting relative displacement of the sub-lattices (Appendix A). This has led to `non-adiabatic' effects, leading
to `phonons behaving badly' \cite{PBB}. A temporal gauge component can also be introduced through the chemical
potential \cite{Sch}. The structure of gauge self-energy in the plane has a characteristic logarithmic singularity,
causing `Kohn anomaly' \cite{KA}. This has been experimentally observed \cite{Sood}, and realized in graphene
through electron-phonon interaction \cite{KG}. 
\paragraph*{}Though lattice excitations in graphene couple to Dirac fermions as effective gauge fields,
they do {\it not} a priory possess kinetic energy or a parity-breaking CS term. Both of these can arise through
the induced effects of the underlying fermions to which the optical phonon couples. In case of graphene, the
fermion mass arises through broken valley degeneracy, resulting in a dispersion of the form: $E^2=\vec p^2v_F^2+M^2v_F^4$
\cite{Neto}, with scaling of mass ($M\rightarrow Mv_F^2$) and coupling strength $g\rightarrow gv_F$ by
Fermi velocity $v_F\ll c$, as compared to standard planar gauge theory \cite{Neto}. At low energies,
where massive fermions are not excited, the gauge field dynamics is obtained by integrating out one of the
minimally coupled fermionic fields, $\psi_s$, from the primary Lagrangian:

\bea
{\cal L}&=&\bar{\psi}_s\left(i\gamma^{\mu}_s\partial_{\mu}-M_sv_F^2-gv_F\gamma^{\mu}a_{\mu}\right)\psi_s\nonumber\\
&+&\bar{\psi}_{s'}\left(i\gamma^{\mu}_{s'}\partial_{\mu}-M_{s'}v_F^2-gv_F\gamma^{\mu}a_{\mu}\right)\psi_{s'},\label{N01}
\eea
with $s=-s'=\pm$ denoting the two valleys $K_\pm$ of graphene, with respective Dirac matrices
$\gamma^\mu_s=(\sigma_3,i\sigma_1,is\sigma_2)$, in momentum-space \cite{Neto}. 
The covariant momentum operator is defined as $i\partial_\mu:=(ic\partial_0,iv_F\vec{\nabla})$.
The corresponding lowest order quantum contribution to the remaining tree-level Lagrangian,
is expressed through the vacuum polarization tensor \cite{IZ}, split into parity even and odd parts as,

\bea
\Pi^{\mu\nu}(q)&=&\Pi_e(q)\left\{q^{\mu}q^{\nu}-\eta^{\mu\nu}q^2\right\}+\Pi_o(q)\epsilon^{\mu\nu\rho}q_{\rho},\nonumber\\
q^\mu&=&cq^0+v_F\vec{q},\nonumber
\eea
with corresponding form factors $\Pi_{e,o}(q)$ respectively. This further represents the linear response
of the system \cite{IZ,Sch}, which is of interest here. The topological part, responsible for the induced
CS term, is unaffected by contributions from fermion sector beyond 1-loop \cite{BDP,Inf,CH,KR}.
\paragraph*{}Following proper regularization (Appendix B), the vacuum polarization form factors are found to be \cite{NA3,Rao},

\begin{align}
\Pi_e(q)&=\frac{g^2v_F^2}{4\pi}\left[\frac{1}{|q|}\left(\frac{1}{4}+\frac{M_s^2v_F^4}{q^2}\right)\log\left(\frac{2\vert M_s\vert v_F^2+\vert q\vert}{2\vert M_s\vert v_F^2-\vert q\vert}\right)\right.\nonumber\\
&\qquad-\left.\frac{\vert M_s\vert v_F^2}{q^2}\right],\nonumber\\
\Pi_o(q)&=-is\frac{M_sv_F^2}{4\pi}\frac{g^2v_F^2}{|q|}\log\left(\frac{2\vert M_s\vert v_F^2+\vert q\vert}{2\vert M_s\vert v_F^2-\vert q\vert}\right).\label{4}
\end{align}
Evidently, both the form factors possess logarithmic singularities at the fermionic two-particle threshold,
$q^2=4M^2v_F^4$, and these expressions are valid {\it below} the same, which is the 
domain of bound-state formation. The even form factor $\Pi_e(q)$ influences wave-function renormalization, while the
odd one, $\Pi_o(q)$, is the 1-loop CS contribution. 
\paragraph*{}The effective Lagrangian now takes the form,
\bea
{\cal L}_e&=&\bar{\psi}_{s'}\left(i\gamma^{\mu}\partial_{\mu}-M_{s'}v_F^2-gv_F\gamma^{\mu}a_{\mu}\right)\psi_{s'}+{\cal L}_g;\nonumber\\
\mathcal{L}_g&=&-\frac{1}{4}f_{\mu\nu}f^{\mu\nu}+\frac{\bar{\mu}}{2}\epsilon^{\mu\nu\rho}a_\mu\partial_\nu a_\rho,\nonumber\\
f_{\mu\nu}&=&\partial_\mu a_\nu-\partial_\nu a_\mu,~~~\hbar=c=1,\label{N02}
\eea
with dynamic CS gauge Lagrangian $\mathcal{L}_g$ and field-strength tensor $f_{\mu\nu}$ for the emergent gauge
field $a_\mu$. It is worth mentioning that $-(1/4)f_{\mu\nu}f^{\mu\nu}$ is a compact way to write the kinetic energy
of the lattice vibrations. $\mathcal{L}_g$ is formed by the vacuum polarization tensor, and ${\cal L}_e$ is normalized modulo
$\Pi_e(q\rightarrow0)=1/\left(16\pi\vert M\vert v_F^2\right)$, leading to the expression,

\be
\frac{\bar{\mu}}{2}=-s\frac{i}{2}\frac{\Pi_o(q\rightarrow0,M_s)}{\Pi_e(q\rightarrow0,M_s)}=-2sM_sv_F^2.\label{N03}
\ee
The remaining fermionic field $\psi_{s'}$ can now be integrated out to obtain 1-loop quantum corrections, through
vacuum polarization, having form-factors exactly like those in Eq. \ref{4}, except $M_s$ replaced by $M_{s'}$. 
\paragraph*{} Both species of Dirac fermion are of {\it same} mass in graphene \cite{Neto}, following the electron
Hamiltonian \cite{LS},

\be
H=\sum_i\left[{\cal E}_AA^\dagger_iA_i+{\cal E}_BB^\dagger_iB_i\right]-t\sum_{\langle i,j\rangle}\left[A^\dagger_iB_j+\text{h.c.}\right].\nonumber
\ee
The individual spin densities ${\cal E}_{A,B}$ of $A,B$ sub-lattices yield the mass
$Mv_F^2=\frac{1}{2}\left({\cal E}_A-{\cal E}_B\right)$ for both, akin to a finite chemical potential, as the
sub-lattices have fixed opposite spins. The corresponding Dirac Hamiltonians for the two valleys are
${\cal H}_s=v_F\left(\sigma_xk_x+s\sigma_yk_y\right)+\sigma_zMv_F^2$ at low energies. This equality of mass
cancels out the individual 1-loop topological contributions \cite{Gra2}, unlike the even parts of
vacuum polarization which add up, yielding the dynamic part of the effective gauge Lagrangian. The topological
exciton manifests only if $M_s>M_{s'}$, with the `lighter' Dirac fermion having sufficiently low two-particle
threshold $2M_{s'}$ to validate Dirac dispersion in graphene. It can be integrated-out to yield a CS contribution
that does not cancel-out with $\bar{\mu}/2$ in Eq. \ref{N03}, as the latter will be logarithmically growing
near the threshold (Eq. \ref{4}). This will finally yield the exciton through 1-loop Schwinger-Dyson equation
(SDE) \cite{SDE}. Recently proposed `penta-graphene' displays intrinsic quasi-direct band gap \cite{PG}, which
is a superposition of {\it two} hexagonal arrays, wherein inducing a second 'independent' mass gap may be
possible.
\paragraph*{}It is worth pointing-out that, a pure CS term, dominant one in low energy limit, is devoid of
dynamics. It represents a parity-breaking four-Fermi interaction between the light fermions, akin to the BCS 
heory of superconductivity \cite{BCS}, where the phonon interactions in the Fr\"ohlich Hamiltonian \cite{Fro},
leads to an effective four-Fermi term. Further, the expression for binding energy resembles that of the
Landau-Ginzburg gap equation \cite{GL}. However, here it is a physical bound state formation, unlike the
spontaneous symmetry breaking in superconductivity.
\begin{figure}
\centering 
\includegraphics[width=3.3 in]{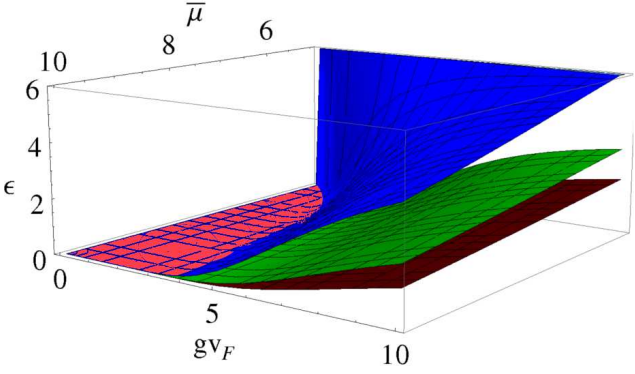}
\caption{Plot of binding energy $\epsilon$ as function of coupling strength $gv_F$ with effective gauge field and the CS coefficient $\bar{\mu}$, for $\vert M_{s'}\vert v_F^2$ being $1$ (maroon), $1.5$ (green) and $2.5$ (blue) respectively (in natural units). All three show suitable regions for bound state formation (small $\epsilon$)
marked in red color, signifying the importance of the condition $\bar{\mu}>2\vert M_{s'}\vert v_F^2$.}\label{f2}
\end{figure}

\paragraph*{}The modification of dynamics at tree-level due to quantum corrections
can be represented through the SDE. Incorporating covariant gauge-fixing, the non-perturbative 1-loop
propagator, containing contributions from an infinite number of bubble diagrams, is obtained as (Appendix B),

\begin{align}
G_F^{\mu\nu}(q)&\equiv\frac{1}{\left[q^2\{1+\Pi_e(q)\}^2+\left\{\Pi_o(q)+i\bar{\mu}\right\}^2\right]q^2}\nonumber\\
&\times\left[\left(q^{\mu}q^{\nu}-\eta^{\mu\nu}q^2\right)\{1+\Pi_e(q)\}\right.\nonumber\\
&\qquad-\left.\epsilon^{\mu\nu\rho}q_{\rho}\left\{\Pi_o(q)+i\bar{\mu}\right\}\right]-\xi\frac{q^{\mu}q^{\nu}}{q^4},\label{12}
\end{align}
with a non-trivial, gauge-invariant pole corresponding to $q^2\{1+\Pi_e(q)\}^2+\left\{\Pi_o(q)+i\bar{\mu}\right\}^2=0$,
which can represent a physical state \cite{IZ}. Here the over-all scaling, introduced in Eq. \ref{N03}, is
factored-out. Such a pole was obtained by Hagen \cite{H}, in `pure' CS QED$_3$
(${\cal L}_g$ replaced with ${\cal L}_{CS}$ in Eq. \ref{N01}, with parameters suitably redefined), just below
the fermionic two-particle threshold: $q^2\approx(2\vert m\vert-\epsilon)^2,~~0<\epsilon\ll1$, interpreted as
a fermion-antifermion {\it spin}-1 bound state (exciton), as planar particle and antiparticle have the same
spin projection \cite{Boy}. It has a `binding energy' that translates as $\epsilon\approx4\vert M_{s'}\vert v_F^2\exp\left(-4\pi\bar{\mu}/g^2v_F^2\right)$ in graphene parameters. In presence of the dynamic
term, we obtain a significantly modified expression:

\be
\epsilon\approx4\vert M_{s'}\vert v_F^2\exp\left\{\frac{4\pi}{g^2v_F^2}\left(2\vert M_{s'}\vert v_F^2-\bar{\mu}\frac{\vert M_{s'}\vert}{M_{s'}}\right)\right\}.\label{14}
\ee
In both these cases, presence of ${\cal L}_{CS}$ at tree-level in the effective theory, is necessary for obtaining a
self-consistent (small magnitude) value of the exciton binding energy, that disappears for vanishing $U(1)$
coupling $g$. As can be checked with different gauge Lagrangians, for $\bar{\mu}=0$, such a value cannot exist, thereby
highlighting the intrinsic topological nature of this exciton.
\paragraph*{}The smallness of the binding energy ($\epsilon$) necessitates a {\it negative} exponent, thereby
fixing a threshold value $\bar{\mu}_t=2\vert M_{s'}\vert v_F^2$ for bound-state formation, highlighting the destabilizing effect
of vacuum fluctuation on the excition due to the kinetic energy term. It further explicates the fact that as
the theory is topologically massive at the tree level (Appendix B), to be realizable in the gauge sector, the exciton
cannot be lighter than the rest mass of the gauge particle. This is one of the main results of this letter. This `non-perturbative'
scenario, even with relatively large coupling strength $gv_F^2$, is justified in terms of the large-N
suppression of higher-order contributions \cite{KR}. For a negative exponent, a shallow bound state is physically
meaningful for {\it small} coupling $gv_F^2$, and deep otherwise, maintaining the self-consistency regarding
$\epsilon\ll \vert M_{s'}\vert v_F^2$. The parametric regions for attaining this exciton is shown in Fig. \ref{f2}.
\paragraph*{}Interestingly, the appearance of $v_F^2$ in the denominator of the exponent in Eq. \ref{14}, makes
the transition about $\bar{\mu}_t$ more prominent, than what it would have been in vacuum, as
$v_F\ll 1$ in natural units. Hence, though both exciton binding energy and formation threshold are
small in graphene, due to smallness of $v_F$, the sensitivity to the bound state formation will be considerably
higher, making experimental verification more likely.
\paragraph*{}The physical properties of this exciton are most evident from the renormalization coefficients, large-N
protected beyond 1-loop, that can be read-off from Eq. \ref{12}. The Lehmann weight $Z_3=\left[1+\Pi_e(q)\right]^{-1}$ vanishes
near the two-particle threshold, owing to the logarithmic singularity, marking emergence of bound-state
as per K\"all\'en-Lehmann spectral representation \cite{IZ}. This further shows vanishing of the
renormalized charge $g_r^2v_F^2=Z_3g^2v_F^2$, representing a charge-less state. The renormalized topological
mass: $\bar{\mu}_r=\left[\bar{\mu}-i\Pi_0(q)\right]Z_3$ leads to $\bar{\mu}_r^2\ge4M_{s'}^2v_F^4$ in the same limit,
marking the threshold. 
\paragraph*{}This topological excitation has unit spin
$\frac{\Gamma}{\vert\Gamma\vert}\approx\frac{M_{s'}}{\vert M_{s'}\vert}$ \cite{Boy}, with,

\bea
\Gamma&:=&-2sM_sv_F^2-s'\frac{M_{s'}}{\vert M_{s'}\vert}\vert M_s\vert v_F^2\log\left(\frac{4\vert M_{s'}\vert v_F^2}{\epsilon}\right)\nonumber\\
&\approx&-s'\frac{M_{s'}}{\vert M_{s'}\vert}\vert M_s\vert v_F^2\log\left(\frac{4\vert M_{s'}\vert v_F^2}{\epsilon}\right),\nonumber
\eea
being the net CS coefficient, near the two-particle threshold of the lighter fermion.
It is being proposed as an ideal candidate for controlled spin-transport, unaffected by local electric
fields for being charge-neutral. Its dynamics can be controlled by an {\it external} magnetic field $\vec{B}$,
allowing for spintronics. The corresponding `external' gauge field ($A_\mu$) couples to the Dirac fermions with
the same strength $gv_F$. The resultant effective gauge action, at low energies, contains an interaction term
$a_\mu\Pi^{\mu\nu}A_\nu$ between the two gauge fields, resolvable as,

\bea
\mathcal{L}^m_I=&-&\frac{1}{2}\left[1+\frac{\Pi_e(q\rightarrow2\vert M_{s'}\vert v_F^2,M_{s'})}{\Pi_e(q\rightarrow0,M_{s})}\right]f_{\mu\nu}F^{\mu\nu}\nonumber\\
&+&\Gamma\epsilon^{\mu\nu\rho}a_\mu F_{\nu\rho},\label{ST1}
\eea
with external field tensor $F^{\mu\nu}\equiv F^{ij}=\epsilon^{ij}B_\perp$. $B_\perp$ is the component of $\vec{B}$
normal to the surface of graphene, that couples to the exciton spin pseudo-scalar in the same direction. This 
is clearly evident in the exciton rest-frame, where the first term of Eq. \ref{ST1}, representing orbital motion,
vanishes ($f_{\mu\nu}=0$). However, the gauge-invariant {\it mixed} CS term \cite{Inf,TopM1,Sgl} in the second,
with corresponding valley contributions adding-up \cite{Neto}, survives. This represents exiton spin coupling with
$B_\perp$, with magnetic moment $2\Gamma a_0$, which is logarithmically large near the two lighter fermion
threshold. The {\it temporal} gauge
field component $a_0$, physically equivalent to a chemical potential \cite{Sch}, is unlike the `spatial' ones
$\left(a_x,a_y\right)$, which are resultants of relative sub-lattice displacements \cite{Ando1} and thus,
are geometric in nature. The non-dynamic mixed CS term inseparably associates quantized magnetic flux to an
electric charge \cite{FA1,FA2}, rendering a non-zero $a_0$ that represents electrostatic potential due to 
charge polarization induced by $B_\perp$ \cite{Boy}. Additionally, the planar spin being a pseudo-scalar,
$\mathcal{L}^m_I$ equivalently represents interaction of exiton spin with external gauge momentum, akin to a
reduced Pauli-Lubanski `pseudo-scalar'.
\paragraph*{}The effect of thermal fluctuations on this topological exciton is demonstrated through a finite
temperature treatment \cite{DasFT}, yielding the temperature dependent extension to $\Pi^{\mu\nu}$ (Appendix C).
Interestingly, the corresponding 1-loop propagator has {\it two} non-trivial poles, with only one 
being physically acceptable. Near two-fermion threshold, the corresponding expression for exciton
binding energy obtained as,

\begin{align}
\epsilon\approx&4\vert M_{s'}\vert v_F^2\exp\left[-T\frac{4\pi}{\vert M_{s'}\vert v_F^2}\Pi_t-8\pi\frac{\bar{\mu}}{g^2v_F^2}\right.\nonumber\\
\qquad &~~~~~~~~~~~~~~~~~~~~~~~~~+\left.16\pi\frac{\vert M_{s'}\vert}{g^2}-1\right],\label{24}
\end{align}
correctly depicting smooth evaporation of the exciton at sufficiently high temperature. In the conventional
high-temperature approximation ($T\gg q,m$), the finite temperature form factor is $\Pi_T(q,T)=Tg^2v_F^2\Pi_t(q)$,
where,

\be
\Pi_t(q)=\frac{1}{2\pi}\log(2)\frac{1-\frac{2}{3}\vert \frac{q}{q_0}\vert^2}{\vert \frac{q}{q_0}\vert-\vert \frac{q}{q_0}\vert^3},\label{23}
\ee
is temperature independent, as per dimensional arguments \cite{DasFT}. In the $T=0$ limit, $\epsilon$ 
goes back to the zero-temperature expression, modulo a constant, owing to non-analytic continuation from
non-covariant $T\neq0$ sector to covariant $T=0$ one \cite{DasFT}. The condition
for exciton melting can be estimated as $T\gg\left(\vert M\vert v_F^2/4\pi\Pi_t\right)$ from this, which
may be physically verifiable, provided the corresponding temperature is still within the low-energy domain
where Dirac dispersion is valid. 
\paragraph*{}Near two-fermion threshold in graphene, one has $\Pi_t\approx(1/4\pi)\log(2)\vert q_0/M_{s'}v_F^2\vert$,
as $v_F\ll1$. The presence of Fermi velocity $v_F$ enhances the temperature effect, making the exciton more
vulnerable to melting in grahphene, than in vacuum. This increases the possibility of experimental observation,
given that the crystal does not melt first.
\paragraph*{}In summary, we have demonstrated that emergent Dirac fermions coupled with phonon-induced gauge
fields, effecting a dynamic CS gauge Lagrangian, yields intrinsically topological novel exciton in planar systems. 
Appearing in the gauge sector, they are charge-neutral spin-1 excitations, ideal for spin-transport in 
graphene-like systems, where massive Dirac modes have been realized. This exciton is characterized by a
parametric threshold, reflecting competition of quantum fluctuations with topological stability. Further, it 
smoothly disappears at high temperatures. The dynamics of this exciton can be controlled 
by an external magnetic field, leading to spin-transport. The observation of the
topological planar exciton will not only demonstrate the gauge-invariant mass of a propagating spin-1 bound state,
but will also substantiate the corresponding topological spin. The charge-neutrality and the topological origin of
spin ensures its stability against charge-dependent forces and both quantum and thermal fluctuations.
\paragraph*{Acknowledgments:} The authors would like to thank Dr. Vivek M. Vyas for valuable inputs. KA is grateful to Prof. Ashok Das
for many useful discussions and suggestions.

\appendix

\section{Emergence of Gauge Fields in Graphene}
The low-energy optical phonons in graphene can be represented by the relative displacement vector of two
sub-lattice atoms $A$ and $B$ as \cite{Ando1},

\be
\vec{u}(\vec{q})=\frac{1}{\sqrt{2}}\left[\vec{u}_A(\vec{q})-\vec{u}_B(\vec{q})\right],\label{4.1}
\ee
which depends on spatial momentum $\vec{q}$. The second quantized version of the above in configuration 
space is,

\be
\vec{u}(\vec{r})=\sum_{g,\vec{q}}\sqrt{\frac{\hbar}{2NM_c\omega_0}}\left(a_{g,\vec{q}}+a^\dagger_{g,-\vec{q}}\right)\vec{e}_g(\vec{q})\exp(i\vec{q}.\vec{r}).\label{4.2}
\ee
with number of unit cells $N$, carbon atom mass $M_c$ and $g=(l,t)$ labeling longitudinal or transverse 
modes. The phonon creation $(a^\dagger_{g,\vec{q}})$ and annihilation $\left(a_{g,\vec{q}}\right)$ operators
correspond to states with definite polarization $\vec{e}_g(\vec{q})$. The electron-phonon interaction
Hamiltonian, for a low-energy nearest-neighbor tight-binding model, can now be expressed as \cite{Ando1},

\be
{\cal H}_{int}=-\sqrt{2}\frac{\beta\gamma}{b^2}\left(u_y(\vec{r})\sigma_x\mp u_x(\vec{r})\sigma_y\right),\label{4.3}
\ee
corresponding to $K_\pm$ valleys respectively. Here, $b=a/\sqrt{3}$ is the equilibrium bond length, $\gamma=\gamma_0\sqrt{3}a/2$
with $\gamma_0$ being the resonance integral between nearest neighbor carbon atoms, and 

\be
\beta=-\frac{d\log(\gamma_0)}{d\log(b)}.\nonumber
\ee
The appearance of Pauli matrices $\sigma_{x,y}$ in Eq. \ref{4.3} and comparison to the effective Dirac Hamiltonian 
${\cal H}=v_F\vec{\sigma}.\vec{k}+\sigma_zMv_F^2$ identifies a $U(1)$ gauge field with spatial components,

\be
\left\{a_x(\vec{r}),a_y(\vec{r})\right\}=-\sqrt{2}\frac{\beta\gamma}{b^2}\left\{u_y(\vec{r}),-u_x(\vec{r})\right\},\label{4.4}
\ee
that couples minimally to the Dirac fermions. The normalization and re-definitions of the emergent gauge field
is done accordingly. 

\section{Topological Exciton with Excitation Threshold}
\begin{figure}
\includegraphics[width=3.3 in]{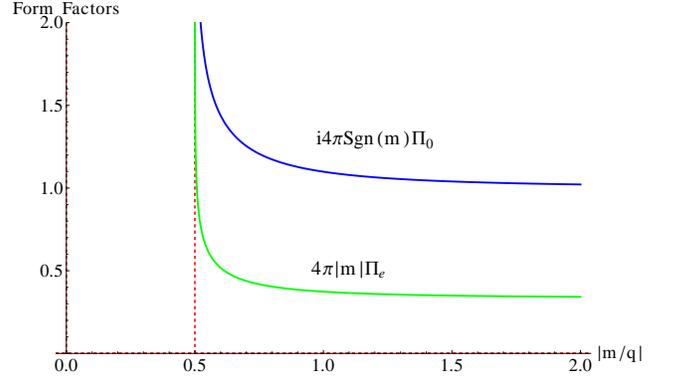}
\caption{Plots depicting regulated amplitudes of even and odd form factors of vacuum polarization tensor. Both have discontinuities,
shown by red dashed lines, that appear at two particle threshold, {\it i.e.}, $\vert m\vert/\vert q\vert=0.5$.
Here $e^2$ is assumed to be $1$, for simplicity, in natural units.}\label{f1}
\end{figure}

While explicating the 1-loop contribution to the gauge propagator, and subsequent emergence of the exciton, we consider standard 
QED$_3$ parametrization for brevity. This amounts to the replacements: $Mv_F^2\rightarrow m$, $gv_F\rightarrow e$ and
$\bar{\mu}\rightarrow\mu$. The role of the tree-level gauge Lagrangian,

\be
\mathcal{L}_g=-\frac{1}{4}f^{\mu\nu}f_{\mu\nu}+\frac{\mu}{2}\epsilon^{\mu\nu\rho}a_{\mu}\partial_{\nu}a_{\rho},\label{A1}
\ee
is to determine the tree-level propagator,

\be
G_{F~0}^{\mu\nu}=-\frac{1}{q^2-\mu^2}\left[\eta^{\mu\nu}-\frac{q^{\mu}q^{\nu}}{q^2}-i\frac{\mu}{q^2}\epsilon^{\mu\nu\rho}q_{\rho}\right]-\xi\frac{q^{\mu}q^{\nu}}{q^4},\label{A2}
\ee
with covariant $R_\xi$ gauge depicted by the last term. It modifies the 1-loop Schwinger-Dyson equation (SDE) \cite{SDE}:

\be
\left[G_F^{\mu\nu}(q)\right]^{-1}=\left[G_F^{(0)~\mu\nu}(q)\right]^{-1}+\Pi^{\mu\nu}(q)-\frac{1}{\xi}q^{\mu}q^{\nu}.\label{5}
\ee
thereby effecting the pole structure of the full propagator $G_F^{\mu\nu}(q)$. The full propagator gets contribution from a series of infinite vacuum
polarization $\left[\Pi^{\mu\nu}(q)\right]$ terms, making the result non-perturbative \cite{IZ}. On integrating
out the fermion field from the full Lagrangian,

\bea
\mathcal{L}&=&\bar{\psi}(x)\left(i\gamma^{\mu}\partial_{\mu}-m\right)\psi(x)\nonumber\\
&-&e\bar{\psi}(x)a^{\mu}(x)\psi(x)\gamma_{\mu}+\mathcal{L}_g,~~~(\hbar=c=1),\label{A3}
\eea
the vacuum polarization contribution to the effective gauge action is obtained as,

\begin{align}
\Pi^{\mu\nu}(q)&=ie^2Tr_D\int\frac{d^3p}{(2\pi)^3}\left[\vphantom{\frac{\partial}{\partial p_{\nu}}}\gamma^{\mu}S_F(p_+)\gamma^{\nu}S_F(p_-)\right.\nonumber\\
&\qquad~~~~~~~~~~~~~~~~~~~~~~~~~~~+ \left.\gamma^{\mu}\frac{\partial}{\partial p_{\nu}}S_F(p)\right];\label{3}\\
S_F(p)&=\left[\gamma^{\mu}p_{\mu}-m\right]^{-1},~~~p_\pm=p\pm\frac{q}{2},~~~p_{\mu}=i\partial_{\mu},\nonumber
\end{align}
with the Schwinger regularization \cite{H} adopted in the second term in the integrand that removes the UV
divergence for zero gauge momentum ($q=0$). The rest of the integral is standard and can be evaluated by 
the derivative expansion method \cite{BDP}, to obtain \cite{NA3,Rao},

\bea
\Pi^{\mu\nu}(q)&\equiv&\Pi^{\mu\nu}_e(q)+\Pi^{\mu\nu}_o(q),\nonumber\\
\Pi^{\mu\nu}_e(q)&=&-\Pi_e(q)Q^{\mu\nu},~~~\Pi^{\mu\nu}_o(q)=\Pi_o(q)\epsilon^{\mu\nu\rho}q_{\rho},\nonumber\\
Q^{\mu\nu}&=&\eta^{\mu\nu}q^2-q^{\mu}q^{\nu},~~~\eta^{\mu\nu}=\text{diag}(1,-1,-1),\nonumber\\
\Pi_e(q)&=&\frac{e^2}{4\pi}\left[\frac{1}{|q|}\left(\frac{1}{4}+\frac{m^2}{q^2}\right)\log\left(\frac{2\vert m\vert+\vert q\vert}{2\vert m\vert-\vert q\vert}\right)-\frac{|m|}{q^2}\right],\nonumber\\
\Pi_o(q)&=&-i\frac{m}{4\pi}\frac{e^2}{|q|}\log\left(\frac{2\vert m\vert+\vert q\vert}{2\vert m\vert-\vert q\vert}\right).\label{4}
\eea
The parity-odd contribution\footnote{In graphene, Dirac matrices: $\gamma^\mu=(\sigma_3,i\sigma_1,is\sigma_2)$
contain the valley index $s=\pm$, which appear multiplied to $\Pi_o$, due to complete antisymmetry of
$\epsilon^{\mu\nu\rho}$.}$\Pi^{\mu\nu}_o(q)$, unique to 2+1 dimensions \cite{Red} arising due to non-zero
trace of three Dirac matrices, is the induced CS contribution. The parity even contribution $\Pi^{\mu\nu}_e(q)$
is responsible for wave-function renormalization \cite{IZ}. The plots of both the form factors
$\left[\Pi_{e,o}(q)\right]$ have been shown in Fig. \ref{f1}, with the well-known singularities at the
two-particle threshold: $q^2=4m^2$. The above results are valid below the same, and 
requires the replacement:

\begin{equation}
\log\left(\frac{2\vert m\vert+\vert q\vert}{2\vert m\vert-\vert q\vert}\right)\rightarrow\log\left(\frac{2\vert m\vert+\vert q\vert}{2\vert m\vert-\vert q\vert}\right)-i\pi,\nonumber
\end{equation}
above it, owing to the corresponding branch-cut. For dynamic CS QED, including the tree level propagator from Eq. \ref{A2},
the full 1-loop propagator is obtained through the SDE as \cite{Rao},

\begin{align}
G_F^{\mu\nu}(q)&\equiv\frac{1}{\left[q^2\{1+\Pi_e(q)\}^2+\left\{\Pi_o(q)+i\mu\right\}^2\right]q^2}\nonumber\\
&\times\left[\left(q^{\mu}q^{\nu}-\eta^{\mu\nu}q^2\right)\{1+\Pi_e(q)\}\right.\nonumber\\
&\qquad-\left.\epsilon^{\mu\nu\rho}q_{\rho}\left\{\Pi_o(q)+i\mu\right\}\right]-\xi\frac{q^{\mu}q^{\nu}}{q^4}.\label{12}
\end{align}
This leads to the non-trivial pole governed by,

\be
q^2\{1+\Pi_e(q)\}^2+\left\{\Pi_o(q)+i\mu\right\}^2=0.\label{13}
\ee
The near two fermion threshold is parametrized as \cite{H}, $q^2=\left(2\vert m\vert-\epsilon\right)$, where $\epsilon$
is small and positive, and to be interpreted as the {\it binding energy} of the exciton. On expanding the form factors
to lowest power of $\epsilon$ and noticing that the logarithm diverges much faster than any other term in their expressions \cite{H},
one obtains,

\bea
\Pi_e(q)&\cong&\frac{1}{4\pi}\left[\frac{1}{2\vert m\vert}\left(\frac{1}{4}+\frac{1}{4}\right)\log\left(\frac{4\vert m\vert}{\epsilon}\right)-\frac{1}{4\vert m\vert}\right]\nonumber\\
&\cong&\frac{1}{16\pi\vert m\vert}\left[\log\left(\frac{4\vert m\vert}{\epsilon}\right)-1\right]~~~\text{and}\nonumber\\
\Pi_o(q)&\cong&-i\frac{1}{8\pi}\frac{m}{\vert m\vert}\log\left(\frac{4\vert m\vert}{\epsilon}\right).\label{29}
\eea
Then, from Eq. \ref{13}, the binding energy of the exciton is,

\be
\epsilon\approx4\vert m\vert\exp\left\{\frac{4\pi}{e^2}\left(2\vert m\vert-\mu\frac{\vert m\vert}{m}\right)\right\}.\label{14}
\ee
As expected, the above expression of binding energy yields Hagen's result as a special case, without the first term in
the exponent.

\section{Effect of Thermal Fluctuation on Exciton}
The extension of above system to finite temperature has been carried out in this section. In 2+1 dimensions,
results were known for mass-less fermion case \cite{Dorey} and induced CS term \cite{BDP} at low energies. We
here obtain the general results, near two-particle threshold. Adopting the real-time formalism of QFT, the
fermionic propagator at $T\neq0$ is obtained from that at $T=0$ as \cite{Re},

\begin{eqnarray}
S_F(p,\beta)&=&U(p,\beta)S_F^0(p)U^T(p,\beta)=S_F^0(p)+S_F^{\beta}(p)\nonumber\\
&\equiv&\frac{1}{\gamma.p-m}+2i\pi n_F(|p^0|)(\gamma.p-m)\delta\left(p^2-m^2\right);\nonumber\\
U(p,\beta)&=&\left(\begin{array}{cc}\cos\theta_\beta & -\sin\theta_\beta \\
\sin\theta_\beta & \cos\theta_\beta \end{array} \right),~~~n_F(|p^0|)=\frac{1}{1+e^{\beta\vert p^0\vert}},\nonumber\\
\theta_\beta&=&\sin^{-1}\left\{\sqrt{n_F(|p^0|)}\right\},~~~\beta=1/T.
\end{eqnarray}
On separating-out the zero-temperature contribution, the finite temperature correction
is identified as,

\begin{align}
\Pi^{\mu\nu}_{\beta}(q)&\equiv iT_r\int_p\left[\gamma^{\mu}S_F^0\left(p_+\right)\gamma^{\nu}S_F^{\beta}\left(p_-\right)+\gamma^{\mu}S_F^{\beta}\left(p_+\right)\right.\nonumber\\
&\qquad\times\left.\gamma^{\nu}S_F^0\left(p_-\right)+\gamma^{\mu}S_F^{\beta}\left(p_+\right)\gamma^{\nu}S_F^{\beta}\left(p_-\right)\vphantom{S_F^0}\right],\nonumber\\
p_\pm&=p\pm\frac{q}{2},~~~p_{\mu}=i\partial_{\mu}.\label{15}
\end{align}
We first proceed to obtain the
contribution to the even part of vacuum polarization, for which we follow the formalism by Weldon \cite{Weldon},
originally carried-out in 3+1 dimensions. Introduction of finite temperature includes the notion of a thermal bath as a physical
reference frame, thus breaking the manifest Lorenz co-variance. However, it is possible 
to obtain a Lorentz covariant formulation by projecting onto and out of the bath coordinate vector
$\{u^{\mu}\}$ \cite{DasFT}. Then the temperature dependent part of the
even component of vacuum polarization can be expressed as,

\bea
\Pi^{\mu\nu}_e(q,T)&=&\Pi_T(q,\omega)P^{\mu\nu}+\Pi_L(q,\omega)R^{\mu\nu},\nonumber\\
P^{\mu\nu}&=&\eta^{\mu\nu}-u^{\mu}u^{\nu}+\frac{\tilde q^{\mu}\tilde q^{\nu}}{Q^2},\nonumber\\
R_{\mu\nu}&=&-\frac{1}{q^2Q^2}(Q^2u_{\mu}+\omega\tilde q_{\mu})(Q^2u_{\nu}+\omega\tilde q_{\nu}),\nonumber\\
\omega&=&q.u,~~~Q^2=\omega^2-q^2,\nonumber\\
\tilde q^{\mu}&=&q^{\mu}-\omega u^{\mu}.\label{16}
\eea
Here $P^{\mu\nu}$ is transverse and $R^{\mu\nu}$ is longitudinal in nature and are orthogonal to
each-other. The {\it thermal} form-factors are expressed as:

\bea
\Pi_L(q,\omega)&=&-\frac{q^2}{Q^2}u_{\mu}u_{\nu}\Pi^{\mu\nu},\nonumber\\
\Pi_T(q,\omega)&=&-\frac{1}{2}\Pi_L(q,\omega)+\frac{1}{2}\eta_{\mu\nu}\Pi^{\mu\nu},\nonumber\\
Re\left(\eta_{\mu\nu}\Pi^{\mu\nu}\right)&=&e^2G_f(q,\omega),\nonumber\\
Re\left(u_{\mu}u_{\nu}\Pi^{\mu\nu}\right)&=&e^2H_f(q,\omega).\label{17}
\eea
The real part of $\Pi^{\mu\nu}$ is of interest here, as the imaginary part of the same corresponds
to the induced CS term, which was obtained by Babu {\it et al} \cite{BDP}, along with the $T=0$ contribution, in
2+1 in the high temperature limit $T\gg q,m$ suitable for present discussion. We will obtain
the general result in the imaginary-time formalism shortly, and show the mentioned result as a
limiting case.
\paragraph*{} From Eqs. \ref{16} and \ref{17}, one obtains:

\begin{eqnarray}
G_f(q,T)&\equiv&\frac{1}{\pi}\sum^1_{s=-1}\int_0^{\infty}\frac{q_0\omega_p+s2m^2}{\vert q\vert\sqrt{4(\omega_p^2-sq_0\omega_p-m^2)+q^2}}\nonumber\\
&\times&\frac{d\omega_p}{e^{\omega_p/T}+1}~~~\text{\&}\nonumber\\
H_f(q,T)&\equiv&\frac{1}{3\pi}\sum^1_{s=-1}\int_0^{\infty}\frac{q_0\omega_p+s(2m^2+\frac{1}{2}q^2-\frac{3}{2}q_0^2)}{\vert q\vert\sqrt{4(\omega_p^2-sq_0\omega_p-m^2)+q^2}}\nonumber\\
&\times&\frac{d\omega_p}{e^{\omega_p/T}+1},\nonumber
\end{eqnarray}
where $\omega_p^2=\vec{p}^2+m^2$. The finite temperature contribution does not introduce additional divergences \cite{DasFT}.
It is well-known that the above $\omega_p$-integrals cannot be solved exactly \cite{DasFT}. As mentioned before, we
ought to take the high temperature (T) limit with $\omega_p:=xT$, finally yielding,

\bea
\Pi_L(q,T)&\approx&-\frac{T}{3\pi}e^2\log(2)\frac{\vert \frac{q}{q_0}\vert}{1-\vert\frac{q}{q_0}\vert^2}~~~\text{and}\nonumber\\
\Pi_T(q,T)&\approx&\frac{T}{2\pi}e^2\log(2)\frac{1-\frac{2}{3}\vert \frac{q}{q_0}\vert^2}{\vert \frac{q}{q_0}\vert-\vert \frac{q}{q_0}\vert^3}.\label{18}
\eea
These contributions are linear in T, as in the hard 
thermal loop (HTL) approximation for 2+1. This is expected from dimensional arguments as the 3+1
counterparts are quadratic in T \cite{TFT}.
\paragraph*{} To obtain the 1-loop gauge propagator at finite temperature,
the obtained finite temperature results are to be molded in suitable forms. By choosing $u=(1,0,0)$, 

\bea
\Pi^{\mu\nu}_e(q,T)&=&\Pi_T\eta^{\mu\nu}+\frac{1}{\vec q^2}\left(\Pi_T-\frac{q_0^2}{q^2}\Pi_L\right)q^{\mu}q^{\nu}\nonumber\\
&-&\frac{q_0}{\vec q^2}(\Pi_T-\Pi_L)\left(q^{\mu}u^{\nu}+u^{\mu}q^{\nu}\right)\nonumber\\
&+&\frac{q^2}{\vec q^2}(\Pi_T-\Pi_L)u^{\mu}u^{\nu},\label{19}
\eea
which is to be substituted in the finite temperature 1-loop SDE,

\begin{figure}
\centering 
\includegraphics[width=3.35 in]{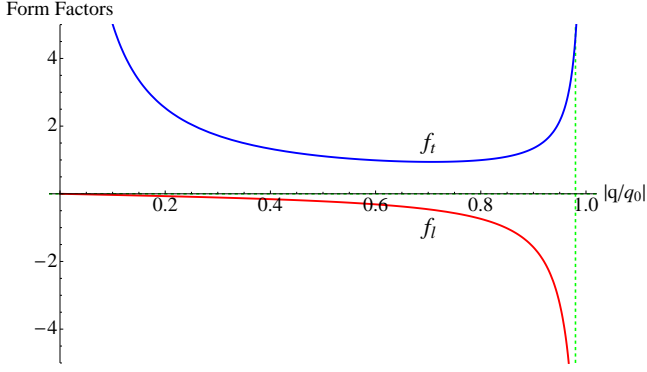}
\caption{Plots of finite-temperature parts of coefficients of vacuum polarization
tensor as functions of $\vert q/q_0\vert$. Here, $f_{l,t}=2\pi\Pi_{l,t}/\log(2)$, the prior can be
large and positive only for unphysical values of $\vert q/q_0\vert$, {i. e.}, larger than
$1$, beyond the dashed green line.}\label{f3}
\end{figure}

\bea
\left\{G_F^{\mu\nu}(q,T)\right\}^{-1}&\equiv&\left\{G^{F~0}_{\mu\nu}(q,T)\right\}^{-1}+\Pi^{\mu\nu}_e(q,0)+\Pi^{\mu\nu}_e(q,T)\nonumber\\
&+&\Pi^{\mu\nu}_0(q,T)+\frac{1}{\xi}q^{\mu}q^{\nu},\nonumber\\
\Pi^{\mu\nu}_o(q,T)&=&-i\frac{me^2}{4\pi}\int_0^1dx\frac{1}{a}\mathcal{I}_m\cot\pi\left(X+is\frac{a}{2\pi T}\right)\nonumber\\
&\times&\epsilon^{\mu\nu\rho}q_{\rho},\label{20}
\eea 
with $a^2=m^2-x(1-x)q^2$ and $X=\frac{1}{2}+xr$, where $r=0,\pm1,\pm2...$
Here, $\Pi^{\mu\nu}_o(q,T)$ is the complete finite temperature odd 1-loop contribution at suitable
high temperature and low momentum limits that falls back to the low-energy expression, given in Ref.
\cite{BDP}, for $q\rightarrow0$. In that domain, up to $\mathcal{O}(1/T)$, the form factor
$\Pi_o(q,T)$ is independent of the external momentum $q$ as has been checked explicitly. The full tree
level gauge propagator at finite temperature (including zero temperature contributions), $G^{F~0}_{\mu\nu}(q,T)$,
is obtained through the replacement \cite{Dic},

\be
-\frac{1}{q^2-m_g^2}\rightarrow-\frac{1}{q^2-m_g^2}-\frac{2\pi}{e^{q_0/T}-1}\delta(q^2-m_g^2),\nonumber
\ee
in the over-all factor. The finite-temperature contribution vanishes for $q^2\neq m_g^2$,
where $m_g$ is the mass of the gauge particle. Since $q$ is the external momentum, and we eventually
are interested in the region just {\it below} the two-particle threshold, this contribution
can be neglected from the onset as either $m_g^2=0$ or $m_g^2=\mu^2<4m^2$, with quantum corrections included. Thus,
one can work with the zero-temperature tree-level propagator, which has been verified directly too.
\paragraph*{}The expressions for temperature-dependent exciton binding energy, $\epsilon$, that
we are going to obtain are expected to yield zero temperature results in the smooth limit $T\rightarrow 0$.  
\paragraph*{}In presence of tree-level propagator, one obtains,

\bea
G^{\mu\nu}(q,T)&\equiv& a(q,T)\eta^{\mu\nu}+b(q,T)q^{\mu}q^{\nu}+c(q,T)u^{\mu}u^{\nu}\nonumber\\
&+&d(q,T)(q^{\mu}u^{\nu}+u^{\mu}q^{\nu})+e(q,T)\epsilon^{\mu\nu\rho}q_{\rho};\nonumber\\
a(q,T)&=&\frac{\Pi_T-q^2\left(\Pi_e+1\right)}{\left[\Pi_T-q^2\left(\Pi_e+1\right)\right]^2+q^2\left(\Pi_o+i\mu\right)^2},\nonumber\\
b(q,T)&=&\frac{\left[\Pi_L-q^2\left(\Pi_e+1\right)\right]+\frac{q_0^2}{\vec q^2}(\Pi_T-\Pi_L)}{q^2\left[q^2\left(\Pi_e+1\right)-\Pi_L\right]}a(q,T)\nonumber\\
&+&\frac{\xi}{q^4},\nonumber\\
c(q,T)&=&-\frac{q^2}{\vec q^2}\frac{(\Pi_T-\Pi_L)}{\left[\Pi_L-q^2\left(\Pi_e+1\right)\right]}a(q,T),\nonumber\\
d(q,T)&=&-\frac{q_0}{q^2}c(q,T),\nonumber\\
e(q,T)&=&-\frac{\left(\Pi_o+i\mu\right)}{\Pi_T-q^2\left(\Pi_e+1\right)}a(q,T),\label{21}
\eea
The gauge dependence retains its covariant $R_{\xi}$ form in $b(q)$. Also,
$c(q)$ and $d(q)$ vanish at zero temperature and the corresponding full propagator is retained.
\paragraph*{}At finite temperature, following Eqs. \ref{21}, there are {\it two} non-trivial pole equations,

\bea
\Pi_L-q^2\left(\Pi_e+1\right)&=&0~~~\text{and}\nonumber\\
\left[\Pi_T-q^2\left(\Pi_e+1\right)\right]^2+q^2\left(\Pi_o+i\mu\right)^2&=&0,\label{22}
\eea
without gauge dependence, ensuring physicality. The conventional high-temperature approximation
($T\gg q,m$), along with the $\epsilon$-expansion near two-fermion threshold yields,

\bea
\Pi_o(q,T)&\approx&\frac{\beta e^2}{i8\pi}m,\nonumber\\
\Pi_e(q,0)&\approx&\frac{e^2}{16\pi\vert m\vert}[\log\left(\frac{4\vert m\vert}{\epsilon}\right)-1].\label{23}
\eea
Consequently, the first pole equation
unacceptably shows {\it increase} of binding energy with temperature, in the physical domain $\vert q\vert\le\vert q_0\vert$,
as illustrated in Fig. \ref{f3} by plotting temperature-independent terms, $\Pi_{l,t}(q)=\Pi_{L,T}(q,T)/Te^2$. 
On the other hand, the {\it second} pole equation in Eqs. \ref{22} correctly represents evaporation of the exciton
at sufficiently high temperature. Near two-fermion threshold, on substituting the expressions in Eqs. \ref{18}
and \ref{23} in this pole equation leads to the exciton binding energy,

\be
\epsilon\approx4\vert m\vert\exp\left[-T\frac{4\pi}{\vert m\vert}\Pi_t-8\pi\frac{\mu}{e^2}+16\pi\frac{\vert m\vert}{e^2}-1\right].
\ee

\end{document}